\documentclass[runningheads]{llncs}
\usepackage{graphicx}
\usepackage{booktabs}
\usepackage{hyperref}
\usepackage{listings}
\usepackage{xcolor}
\usepackage{acronym}
\usepackage{comment}
\usepackage{amsmath}

\acrodef{CSP}[CSP]{Cloud Service Provider}
\acrodef{CSPs}[CSPs]{Cloud Service Providers}
\acrodef{CSCs}[CSCs]{Cloud Service Certifications}
\acrodef{CSC}[CSC]{Cloud Service Certification}
\acrodef{NLP}[NLP]{Natural Language Processing}
\acrodef{NL}[NL]{Natural Language}
\acrodef{QA}[QA]{Question Answering}
\acrodef{AMOE}[AMOE]{Assessment and Management of Organisational Evidence}
\acrodef{IR}[IR]{Information Retrieval}
\acrodef{IaaS}[IaaS]{Infrastructure-as-a-Service}
\acrodef{SaaS}[SaaS]{Software-as-a-Service}
\acrodef{BERT}[BERT]{Bidirectional Encoder Representations from Transformers}
\acrodef{RoBERTa}[RoBERTa]{A Robustly Optimized BERT Pretraining Approach}
\acrodef{SQuAD2}[SQuAD2]{Stanford Question Answering Dataset}
\acrodef{EUCS}[EUCS]{European Cybersecurity Certification Scheme for Cloud Services}
\acrodef{NIST}[NIST]{National Institute of Standards and Technology}
\acrodef{CCM}[CCM]{Cloud Control Matrix}
\acrodef{CSA}[CSA]{Cloud Security Alliance}
\acrodef{BiLSTM}[BiLSTM]{Bi-directional Long Short Term Memory}
\acrodef{NLTK}[NLTK]{Natural Language Toolkit}

\definecolor{codegreen}{rgb}{0,0.6,0}
\definecolor{codegray}{rgb}{0.5,0.5,0.5}
\definecolor{codepurple}{rgb}{0.58,0,0.82}
\definecolor{backcolour}{rgb}{0.95,0.95,0.92}

\lstdefinestyle{mystyle}{
    backgroundcolor=\color{backcolour},   
    commentstyle=\color{codegreen},
    keywordstyle=\color{magenta},
    numberstyle=\tiny\color{codegray},
    stringstyle=\color{codepurple},
    basicstyle=\ttfamily\footnotesize,
    breakatwhitespace=false,         
    breaklines=true,                 
    captionpos=b,                    
    keepspaces=true,                 
    numbers=left,                    
    numbersep=5pt,                  
    showspaces=false,                
    showstringspaces=false,
    showtabs=false,                  
    tabsize=2
}

\begin{document}
\title{AMOE: a Tool to Automatically Extract and Assess Organizational Evidence for Continuous Cloud Audit\thanks{This work has received funding from the European Union’s Horizon 2020 research and innovation programme under grant agreement No: 952633.\\This work was partially supported by project SERICS (PE00000014) under the MUR National Recovery and Resilience Plan funded by the European Union - NextGenerationEU.\\The final publication is available at Springer via \url{https://doi.org/10.1007/978-3-031-37586-6_22}.}}
\titlerunning{AMOE: a Tool to Extract and Assess Organizational Evidence}
%
\author{Franz Deimling\inst{1}\orcidID{0009-0006-6285-9373}
\and
Michela Fazzolari\inst{2}%
}
\authorrunning{Deimling et al.}
%
\institute{Fabasoft R\&D GmbH, 4020 Linz, Austria \email{franz.deimling@fabasoft.com}
\and IIT-CNR, 56122 Pisa, Italy \email{m.fazzolari@iit.cnr.it}
}

\maketitle
\begin{abstract}
The recent spread of cloud services has enabled many companies to take advantage of them. Nevertheless, the main concern about the adoption of cloud services remains the lack of transparency perceived by customers regarding security and privacy. To overcome this issue, \ac{CSCs} have emerged as an effective solution to increase the level of trust in cloud services, possibly based on continuous auditing to monitor and evaluate the security of cloud services on an ongoing basis. Continuous auditing can be easily implemented for technical aspects, while organizational aspects can be challenging due to their generic nature and varying policies between service providers. 

In this paper, we propose an approach to facilitate the automatic assessment of organizational evidence, such as that extracted from security policy documents. The evidence extraction process is based on \ac{NLP} techniques, in particular on \ac{QA}. The implemented prototype provides promising results on an annotated dataset, since it is capable to retrieve the correct answer for more than half of the tested metrics. This prototype can be helpful for \ac{CSPs} to automate the auditing of textual policy documents and to help in reducing the time required by auditors to check policy documents.

\end{abstract}

\section{Introduction}
\label{sec:intro}
In recent years, cloud computing has emerged as a widespread solution for businesses seeking cost-effective and scalable IT infrastructure. The advantages brought by cloud services, such as flexibility, cost-efficiency and maintenance reduction, made them an attractive option for companies of all sizes. 

Nevertheless, the adoption of cloud services also implies moving from direct control and governance of data and applications to an indirect form of control. Thus, several concerns have been raised about security, privacy, transparency and trustworthiness. To address these concerns, \ac{CSCs} have become an effective solution to increase the level of trust in cloud services~\cite{lins2018trust}. \ac{CSCs} provide assurance that cloud services are compliant with industry standards and best practices, and that appropriate security controls are in place to protect sensitive data and applications. The goal is to obtain a certificate that proves compliance with one or more security schemes, allowing users and customers to trust the cloud service.

While traditional security certification approaches rely on periodic audits (annually or biannually), continuous auditing has emerged as a method to monitor and evaluate the security of cloud services on an ongoing basis~\cite{windhorst2013dynamic,cabuk2019thetransformation}. Continuous auditing involves the collection, analysis, and evaluation of security-related data and events to provide real-time insights into the security posture of cloud services. Continuous auditing is particularly effective when applied to analyze technical/performance aspects, which can be represented with unambiguous and measurable objectives. 
On the contrary, when considering administrative/organisational aspects, they are generally not analysed, or their analysis is conducted manually. This is due to the fact that organisational aspects, by their nature, are generic, typically expressed by policies written in natural language and varying from one service provider to another. For these reasons, they are difficult to be transformed into objective quantities that can be measured and evaluated automatically. However, they play a critical role in establishing a strong security culture and ensuring compliance with security standards.

Given the challenge mentioned above, the novelty of this work resides in the proposal of a tool, namely \ac{AMOE}, to enable evidence extraction and assessment of policy documents containing organizational aspects. This tool has been developed within the context of the MEDINA project~\cite{orue2021medina}, which aims at providing a security framework for continuous audit-based certification in compliance with cloud security certification schemes, in particular with the \ac{EUCS}. These schemes, indeed, include both technical and organisational controls. The latter are not suitable to be automatically monitored like technical requirements, but they still need to be verified and checked in a way that enables the continuous and automatic issuing of security certificates. 

The \ac{AMOE} tool relies on a set of organisational metrics specifically defined for this purpose, which aim to measure concrete parts in the policy documents. The proposed approach is based on \ac{NLP} techniques since policy documents contain organizational evidence expressed in \ac{NL}. The audit information is then extracted from such texts, exploiting a \ac{QA} system. Indeed, question answering has been successfully used in various information retrieval tasks, e.g. to build a search system for COVID-19-research documents~\cite{esteva2020cosearch}.

To evaluate the suitability of our proposal, a set of retrieval pipelines has been tested utilizing \ac{QA} to extract the relevant audit information.
Then, the different evidence extraction pipelines have been tested by conducting a set of experiments, which show that the \ac{QA} approach, combined with keyword-based paragraph filtering, leads to promising results.

The proposed tool can be useful for \ac{CSP} compliance managers or auditors to verify the compliance of a policy document with respect to a set of organisational metrics.

This paper is structured as follows: Section~\ref{sec:related} discusses the related work in the domain of security transparency and audit in cloud computing. Section~\ref{sec:background} provides some background information about techniques and methods used in this work.
The data used for the experimentation are described in Section~\ref{sec:data}. Section~\ref{sec:methodology} describes the methodology used in the approach proposed in this paper. 
The experimental results and a discussion are provided in Section~\ref{sec:results}.
Finally, Section~\ref{sec:conclusion} concludes the paper and suggests some improvements that can be developed.

\section{Related work}\label{sec:related}
Several contributions have been proposed in the last decade dealing with continuous auditing and dynamic certification of cloud services. 
In 2013, Cimato et al.~\cite{cimato2013towards} introduced the concept of cloud service certification and they drew an initial proposal containing a conceptual framework with the specification of certification models. Afterwards, Anisetti et al.~\cite{anisetti2015toward,anisetti2016acertification} proposed a test-based certification framework for automating the certification process and a cloud engineering methodology based on it. 

The authors in~\cite{stephanow2015towards} focus on \ac{IaaS} providers and propose an approach that aims to continuously detect ongoing changes in the services. To assess the impact of these changes on customer requirements, low-level metrics are used, which are derived from widely deployed \ac{IaaS} components. Furthermore, the authors show how these low-level metrics can be used to construct complex metrics to support the validation of generic requirements. A further approach proposed in~\cite{stephanow2017towards} focuses on security certification for \ac{SaaS} providers. In this work, the authors develop a method to support continuous certification of \ac{SaaS} applications, using web application testing techniques. In particular, they exploit SQLMap, a tool for web application testing.

One of the first attempts to summarize contributions in this field can be found in the paper published by Lins et al.~\cite{lins2015what}, where the authors conducted a systematic literature review to identify automated auditing methods that could be applied within the context of cloud computing. Then, they propose a conceptual architecture for continuous auditing of cloud services, identify the cloud services criteria that should be continuously audited, evaluate applicable methodologies, and highlight the components and processes required for successful implementation~\cite{lins2018trust}. 

Work on the subject of continuous auditing has increased in recent years. In~\cite{knoblauch2019reducing} the authors highlight the difficulties in the implementation of continuous auditing due to the lack of standardized approaches.
In this regard, they propose a solution by defining a standardized way of establishing a continuous audit process and by providing a methodology to realize it, by leveraging an Audit API. 
The work in~\cite{kunz2017aprocess} discusses the challenges of integrating continuous certification checks in existing certification processes for cloud services. The authors analyze and generalize traditional certification processes and then propose a novel certification process model to address these challenges and support continuous certification techniques.

The work in~\cite{torkura2021continuous} proposes a novel cloud security system, namely CSBAuditor, to continuously monitor and audit cloud infrastructure. CSBAuditor uses the reconciler pattern and state transition analysis to detect misconfigurations, malicious activities, and unauthorized changes. Furthermore, a new scoring system is proposed, i.e. Cloud Security Scoring System, which uses security metrics to compute severity scores for detected vulnerabilities. The authors also highlight the limitations of existing security management processes and the need for more customer-centric security mechanisms to protect cloud infrastructure.

While this is a broad topic and they do not provide technical implementation, research teams such as Banse et al.~\cite{banse2021cloud} focus on specific aspects such as static code analysis, which can be part of the evidence gathering needed for continuous audits. Some of the theoretical concepts are implemented in different technical evidence gathering tools (e.g. Clouditor\footnote{https://github.com/clouditor/clouditor}) and have been tested in audit use cases. However, we are interested in the technical feasibility of handling and integrating organisational evidence in the automated audit process.

\section{Background}\label{sec:background}
Audits of cloud computing services look for evidence that a cloud provider is using best practices, complies with appropriate standards and meets certain benchmarks in delivering its services. Thus, when performing a cloud audit, one of the first steps involves evidence collection, i.e. gathering, integrating and processing logs, policies and metadata coming from the cloud.

Security certification schemes such as the \ac{EUCS} contain several objectives to define technical and organizational measures to ensure the security of the systems\footnote{https://www.enisa.europa.eu/publications/eucs-cloud-service-scheme}. There are requirements that call for policies and procedures and provide guidelines for the content of the policies. Similarly, \acs{NIST} classify some controls (similar to requirements in \ac{EUCS}) to address administrative, technical and physical aspects\footnote{https://nvlpubs.nist.gov/nistpubs/SpecialPublications/NIST.SP.800-53r5.pdf}. The Continuous Audit Metrics Catalog\footnote{https://cloudsecurityalliance.org/artifacts/the-continuous-audit-metrics-catalog/} released by \ac{CSA} lists metrics to deal with technical security measures. The authors make a distinction between the evaluation of technical measures and the evaluation of policy and procedural controls. They argue, that the technical ones can be managed by automated techniques to collect evidence to prove their effectiveness and that this idea can be extended partially to policy and procedural controls. 
Independent of the security scheme vendor, there is a distinction between technical and non-technical (administrative or organizational) measures that are specified by the security controls or requirements. However, the security control descriptions are too generic to classify them as organizational or technical. Most of them cover both technical and organizational aspects and call for automation if possible, especially for higher levels of assurance.

When considering the organizational part of a requirement, the evidence collected for the cloud audit is mostly available in textual form, thus a possible way to process it is to use \ac{IR} approaches, which has shown increasingly accurate results over the last years. In particular, \ac{NLP} techniques can help retrieve accurate results from large amounts of text. 
In this section, we provide some background information regarding \ac{NLP} that has been used in the creation of the \ac{AMOE} tool.

\subsection{Text Pre-Processing}
We exploited several text pre-processing techniques to clean the text analyzed by the \ac{AMOE} tool.

\textbf{Tokenization} is the first step in any \ac{NLP} pipeline and is the process of breaking a stream of textual data into words, terms, sentences, symbols, or some other meaningful elements called tokens~\cite{webster1992tokenization}. A tokenizer breaks unstructured data and natural language text into chunks of information that can be considered as discrete elements.

\textbf{Lemmatization}, in contrast to stemming, does not remove the suffixes of words but tries to find the dictionary form of a word on the basis of vocabulary and morphological analysis of a word~\cite{schutze2008introduction,balakrishnan2014stemming}. Furthermore, it could help to match synonyms such as “hot” and “warm” – or “car”, “cars” and “automobile”~\cite{balakrishnan2014stemming}.

\textbf{Stop word removal} involves removing common words that are irrelevant for a query, as they occur with the same likelihood over all documents. In fact, words such as “the”, “if”, “but”, “and”, etc. do not add additional content to the document. Removing such words can improve retrieval performance~\cite{wilbur1992automatic}. However, for some queries, not removing the stop words leads to more precise results~\cite{schutze2008introduction}.

\subsection{Language Encoding Models}
In the last decade \ac{IR} research has brought us new ways to approach \ac{NLP} tasks by developing Language Encoding Models, i.e. models designed to represent words or text with the goal of capturing their underlying meaning or semantic information.

An example of Language Encoding Model is the \ac{BERT} model, which was introduced by Devlin et al.~\cite{devlin2018bert}. This model has been successfully applied over the last few years and additional research has been built upon it. A resulting refinement is \ac{RoBERTa}~\cite{DBLP:journals/corr/abs-1907-11692}, which manages to achieve very good performance on different datasets. We would like to mention one particular dataset in the following, e.g. \ac{SQuAD2}~\cite{DBLP:journals/corr/RajpurkarZLL16}, as it has been used to fine-tune the \ac{QA} model\footnote{https://huggingface.co/deepset/roberta-base-squad2} applied in the evidence extraction pipelines described in Section~\ref{sec:evidence_extraction}. This dataset is based on a set of Wikipedia articles and annotated answers to 100,000+ questions.
Further models based on FastText~\cite{DBLP:journals/corr/JoulinGBDJM16} for learning word vectors of languages have been presented by Grave et al. in~\cite{DBLP:journals/corr/abs-1802-06893}. These models are trained with data from Wikipedia and Common Crawl  and evaluated on word analogy tasks. For a triplet of words $A : B :: C$, the goal is to guess D. For example, for the triplet $Paris : France :: Berlin : ?$, the answer would be \textit{Germany}. The resulting models can produce word vectors, which can be used for further applications, e.g. text classification or text similarity tasks.

\subsection{\acf{QA}}
\acf{QA} is a task of \ac{NLP} which consists in retrieving one or more answers to a question, given an input text as \textit{context}. Once trained, \ac{QA} models do not require a complex setup to be used. Some \ac{QA} models can work even without context. These systems are especially used in customer service chat-bots, evidence extraction or in semantic search engines~\cite{esteva2020cosearch}.  The use of large numbers of documents increases the probability of including the correct answer, but the \ac{QA} system’s performance can be negatively affected. Therefore, Lee et al.~\cite{DBLP:journals/corr/abs-1810-00494} investigate paragraph ranking to improve answer recall in \ac{QA}. To this aim, they introduce a Paragraph Ranker to increase answer recall and reduce noise. For a set of $N$ documents containing $K$ paragraphs on average, the system only selects the $M$ top-ranked paragraphs. In this way, $N$ is increased while the number of read paragraphs $M$ is reduced, leading to higher answer recall.
Lee et al. calculate the probability that a paragraph contains the answer to the question by using different similarity functions.
This probability is then used in the loss function of the model. 

We used an approach similar to the Paragraph Ranker described by Lee et al.~\cite{DBLP:journals/corr/abs-1810-00494} in the similarity-based evidence extraction method, as described in Section~\ref{subsec:simililarity}. The main difference is that instead of training a model to refine the ranking, the ranking of results for evidence extraction is done directly using the similarity score.
The Paragraph Ranker approach is not applicable in our case because training a model given the two documents, it would likely over-fit.

To this aim, cosine similarity can be used to determine the similarity between documents and texts of different lengths. In fact, despite having different lengths, two documents might be similar. This happens because the relative term frequency could be similar, thus given similar term frequencies, it can be assumed that the content of the documents is similar as well. The denominator of the equation normalizes the length of the vectors~\cite{schutze2008introduction}. The cosine similarity is defined as shown in equation~\ref{eq:eq2}, where $A$ and $B$ could be a document or word vector representations.

\begin{equation}\label{eq:eq2}
cos\ \theta\ =\ \frac{A\cdot B}{|A|\left|B\right|}
\end{equation}

\section{Dataset}
\label{sec:data}
The \ac{AMOE} tool uses two different types of textual data. The first type of data is used for extracting evidence and it is represented by a set of organisational metrics. This set was specifically designed within the MEDINA project. Nevertheless, it includes metrics of general nature and can therefore easily be reused within other frameworks or extended by incorporating new metrics. The second type of data contains the policy documents of \ac{CSP}s.

In the following, the two kinds of data will be described in detail.

\subsection{Organisational metrics}
\label{subsec:org_metrics}

The organisational metrics are quantitative measures that can be associated with organisational requirements, with the aim of defining how compliance can be automatically assessed. In \ac{AMOE} they represent the structured form of a query for the evidence to be assessed. Specifically, a metric consists of a set of attributes such as a name, a description, and some keywords found in the paragraph or section heading of the expected evidence to extract. The metric also includes values useful for the assessment, e.g. an operator, a target value and a data type. An example of metric and its attributes is reported in Table~\ref{tab_metric}.

\begin{table}[!bht]
\begin{center}

\caption{Example of organisational  metric.}\label{tab_metric}%
\begin{tabular}{ll}
\toprule
\textbf{Attribute} & \textbf{Value}\\
\midrule
name & PasswordPolicyQ2\\
description & What is the password’s maximum age according to the password policy?\\
keywords & password, age, maximum\\
operator & $\leq$\\
target value & 100\\
data type & Integer\\
\bottomrule
\end{tabular}

\end{center}
\end{table}

Security certification schemes such \ac{EUCS} aim at enhancing trust in cloud services by defining a set of security requirements. Within this context, metrics are linked to security requirements, thus to assess compliance with a requirement, we need to assess the corresponding metrics. 

The description of each organisational metric is targeted to measure a specific part of a requirement. As the requirements are written in a generic way and allow some broader interpretation, multiple metrics are required to reflect the compliance status of a cloud service or policy to a requirement.

Therefore, the organisational metric description is formulated as a precise question targeted in consequence to measure and assess compliance. This way, the query is human-readable as well as usable by the evidence extraction pipeline. The keywords supplied in the metric can be used to reduce the search space for the evidence in the different approaches explained in Section~\ref{sec:evidence_extraction}.

\subsection{Policy documents}
\label{subsec:policy_docs}
The policy documents are usually formatted as unstructured text in the form of Word documents or PDFs. In this case, the experiments are conducted using the documents supplied by two industry partners. 

Since the documents contain text in natural language, they need to be pre-processed and prepared for evidence extraction. Moreover, they need to be annotated, in order to gain some insights into the performance of the whole system. For this, the INCEpTION software has been chosen~\cite{klie2018theinception}, which is described as “A semantic annotation platform offering intelligent assistance and knowledge management”. Thus, for a set of metrics, the expected evidence (i.e. the answer to the metric description question) is marked directly in the text. The annotations can be extracted as TSV (Tab-Separated Value) files and thus conveniently used for quality checks of \ac{AMOE}. 

The policy documents provided contain multiple policies, so they are addressing different requirements of the \ac{EUCS}. Some documents contain more than 50 pages and this affects the processing duration as well as extraction accuracy. In fact, \ac{AMOE} tool does not know apriori what policies are included in a document, thus every metric needs to be checked, even if it could not be answered since it is not covered by the document.

\section{Methodology} \label{sec:methodology}
In this section, we describe the methodology used within the \ac{AMOE} tool to extract evidence based on metrics from certain documents linked to the organisational requirements and to process and prepare them for assessment. 

\subsection{Pre-Processing}\label{sec:pre_proc}
Both the policy documents and the metrics data need to be pre-processed to be used within the \ac{AMOE} tool.

\subsubsection{Policy Document Pre-Processing}
\begin{figure}
\includegraphics[width=\textwidth]{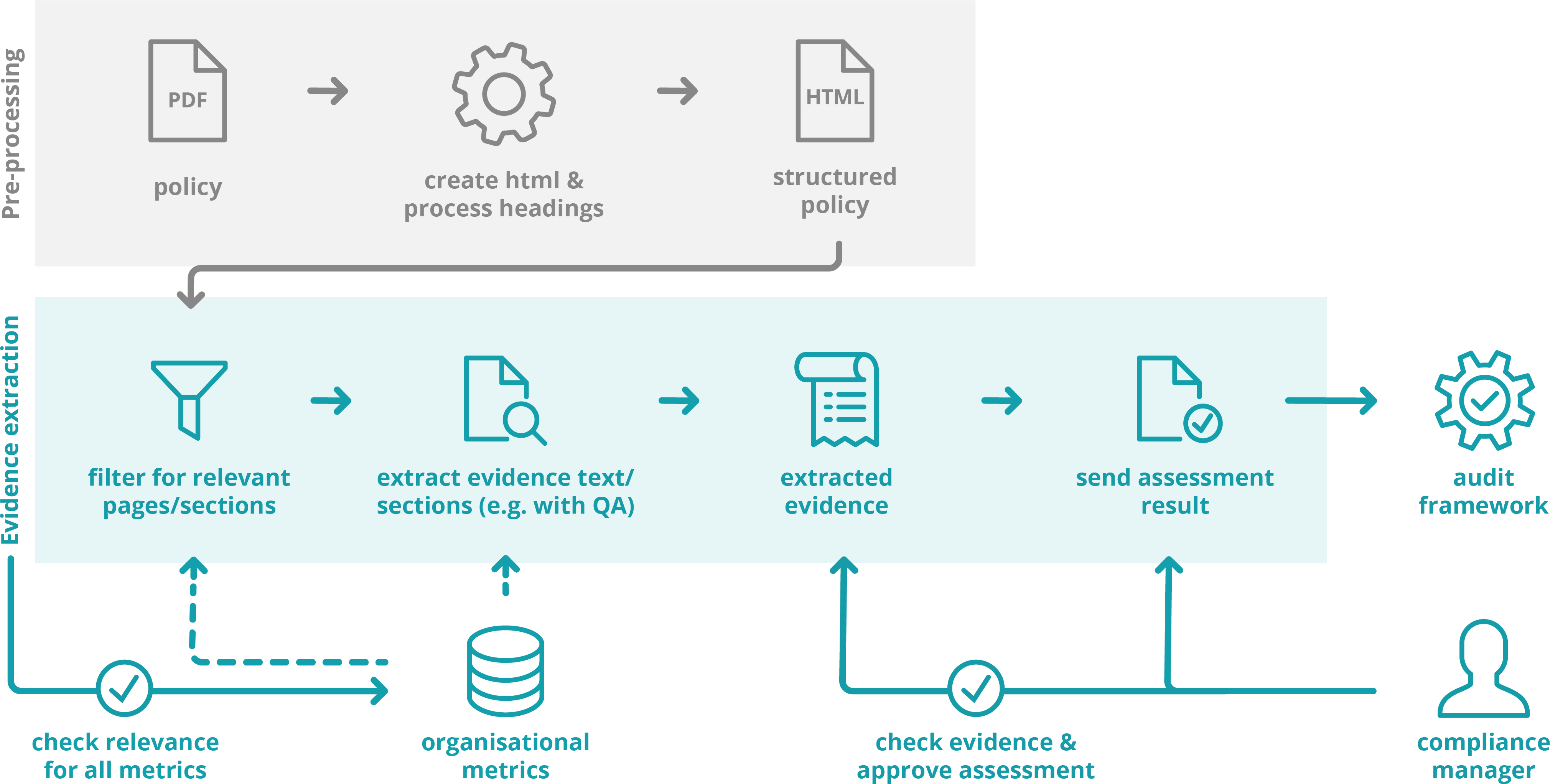}
\caption{Simplified overview of the AMOE evidence pipeline.} \label{fig_overview}
\end{figure}

As already introduced, policy documents are provided in the form of PDF documents. To retain some of the structure given by e.g. section headings, these documents are transformed into HTML. Depending on the document's origin, some headings need further recognition, e.g. through rule-based approaches. This process is depicted in the upper part of Figure~\ref{fig_overview}.

The PDF document is transformed into an HTML with the help of the Poppler library\footnote{https://poppler.freedesktop.org/}, specifically by using the \textit{pdftohtml} tool.  While the document is processed, common errors for section headings are fixed. For example, it can occur that headings are not correctly detected by the initial conversion to HTML. Therefore, the heading tags are merged into a single tag, if the heading spans over multiple lines, or they are added if they have been missed. Short text spans are considered to be a heading if the text size is high (relative to the rest of the document) or if the font is bold.

The result of this process is a structured document containing headings and sections/paragraphs, that eases the information filtering task. In the last stage of the pre-processing, information such as the table of contents, or parts of the headings or footer are removed. The whole process depends heavily on the quality of the PDF, which is reflected in the final HTML document.

Transforming the document to HTML has at least three benefits: 1) it brings structure into an otherwise unstructured text (as seen by the program), 2) the obtained structure can be used to perform better analyses and to infer why something was found or could not be obtained and 3) the document in HTML format can be directly used to present the answers in a User Interface, without additional processing.

\subsubsection{Metric pre-processing}
\label{subsec:metric_pre_proc}
As for the policy documents, the metrics data need to be pre-processed as well. For example, the metric keywords can contain irrelevant words for queries and therefore they are removed. As a basis for the removal, common English stop words defined in the \ac{NLTK} corpus have been used~\footnote{https://www.nltk.org/}. Furthermore, the remaining words are lemmatized and used in the different extraction processes.

\subsection{Evidence extraction}\label{sec:evidence_extraction}
After pre-processing the document, the evidence can be extracted using different
approaches. In the following, the basic approach is first explained, which consists in taking the entire policy document as input and feeding it to the \ac{QA} model. 
However, as previously mentioned, processing very long documents may lead to a deterioration in performance. Therefore, alternative approaches to mitigate this problem are also described. In particular, in some approaches, the documents have been filtered in a preliminary step, in order to reduce the search space for the evidence. 

The bottom part of Figure~\ref{fig_overview} shows the pipeline steps for evidence extraction and connection to a hypothetical component that will use the assessment results somehow. For all the approaches, the extraction of evidence is based on a pre-trained \ac{QA} model, specifically, the \ac{RoBERTa} base model, fine-tuned using the SQuAD2 dataset (roberta-base-squad2~\footnote{https://huggingface.co/deepset/roberta-base-squad2}).

The evidence extracted by the \ac{QA} model is then analyzed. If a target value has been set for a metric, the extracted evidence is translated into a similar datatype (if possible) and an assessment hint is computed by checking the output against the target value with the defined metric operator. Moreover, the \ac{QA} model provides a score that could aid in determining whether the output is relevant since not all queries produce relevant outputs.

\subsection{Baseline: Whole-doc Approach}
\label{subsec:whole_doc}
\begin{figure}[hbt!]
    \centering
    \includegraphics[width=0.9\textwidth]{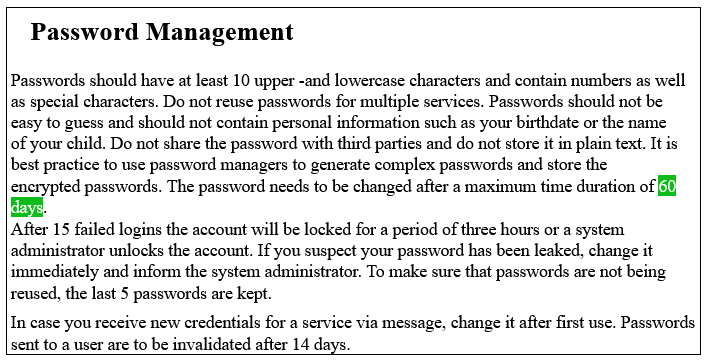}
    \caption{Excerpt of a policy document.}
    \label{fig:policy_doc}
\end{figure}

In the baseline approach, the text of the whole document is fed into the \ac{QA} model to produce a single answer. In the following, we provide an example of how the extraction pipeline works for the metric shown in Table~\ref{tab_metric}. In Figure~\ref{fig:policy_doc} an excerpt of a policy document is reported, which represents the context for the question. 
Table~\ref{tab:evidence_extraction}, instead, shows the question given as input to the \ac{QA} model, which corresponds to the metric description, and the answer (marked in bold) as extracted from the evidence.

\begin{table}[thb!]
    \centering
    \caption{Evidence extraction example.}
    \begin{tabular}{l}
    \toprule
    \textbf{Query}     \\
    What is the password’s maximum age according to the password policy? \\
    \midrule
    \textbf{Evidence/Answer} \\
    The password needs to be changed after a maximum time duration of \textbf{60 days}. \\
    \bottomrule
    \end{tabular}
    \label{tab:evidence_extraction}
\end{table}

The assessment hint is generated by converting the answer from string to integer and by comparing this value with the metric target value, using the metric operator. Thus, for the example reported, the assessment hint will appear as follow:
 \begin{equation*}
     60 \leq 100 \rightarrow True
 \end{equation*}

An (internal) auditor or compliance manager can then approve the assessment result and initiate the workflow needed to obtain a certificate with the help of an audit framework.

The approach described so far can retrieve relevant evidence from small documents containing only a few policies. However, for longer documents, the extraction quality suffers and the computation time is higher (as reported in Section~\ref{sec:results}).
In the following, we describe alternative evidence extraction approaches developed to improve performances.

\subsubsection{Keyword-Based Approach}
\label{subsec:keyword}

The keyword-based approach tries to mitigate the negative effects of document length on performance. In fact, the documents can contain large amounts of text and multiple policies irrelevant to the metric, thus this information needs to be filtered out. The underlying idea consists in mimicking what a human auditor would do. First, search for the relevant section based on some keywords (derived from experience, similar policy documents, etc.), then read the section and decide whether the necessary information is included.
Therefore, relevant sections are sought according to the metric keywords.
If no sections can be found, the whole document is used. Stop words are removed from the metric keywords and the remaining words are lemmatized. 

For this simple approach, a section is considered relevant if the lemmatized section heading words intersect with the metric keywords. Once the filtering is complete, the relevant sections are concatenated into a single input text. Lastly, the pre-trained \ac{QA} model is fed with both the concatenated text as well as the metric question, and a single answer is retrieved. The process then continues as the baseline approach.

The main benefit of the keyword-based approach is that it is quite fast since the model is used only once per metric. However, some evidence could be missed, because sections are excluded based on the keywords. This error occurs if the evidence is in the document, just not in a section with a relevant heading. Therefore, in some cases, the results of this approach depend more on the keywords, than on the \ac{QA} model’s performance itself.

\subsubsection{Score-based approach}
\label{subsec:score}
To deal with the problem arising from the length of the text, one possibility is to consider individual sections instead of the entire document. Thus, in the score-based approach, multiple queries are presented to the pre-trained \ac{QA} model, one for each section of the document. For each query, the context is the text of a section and the question is the metric description. With such an approach, the model provides not one, but as many answers as there are sections. As described in Section~\ref{sec:evidence_extraction}, the \ac{QA} model also returns a score associated with each answer: the higher the probability that an answer correlates with a question, the higher the score. 

Therefore, once all answers and scores are retrieved, the answer with the highest score is returned. This approach stems from the assumption that the section that produces the highest score would be the correct one. However, empirical results and also quantitative comparisons to the annotated samples showed that this is not always the case. This could be probably mitigated by fine-tuning the \ac{QA} model with respect to the domain.

\subsubsection{Similarity-based approach}
\label{subsec:simililarity}
The similarity-based approach was designed to make better use of the metric keywords. To this aim, the metric keywords are lemmatized and transformed into a feature vector by applying the FastText model ($qry\_features$), computed at sentence level instead than at word level. The same is done for each section in the document, and the feature vectors for sections are obtained ($section\_features$). At this point, the cosine similarity measure is computed between the $qry\_features$ and the $section\_features$ of each section. Finally, the pre-trained \ac{QA} model is applied, considering each section as the context and the metric description as the question. Similar to the score-based approach, multiple answers are retrieved but this time the answer corresponding to the section with the highest cosine similarity is returned. This approach is highly dependent on the metric keywords.

\subsubsection{Similarity + score-based approach}
\label{subsec:sim_score}
This approach combines the similarity-based and score-based approaches and it works in the same way. This time, for each answer, both scores and cosine similarities are computed and added up. Finally, the answer with the highest sum is returned, with the idea of balancing out the flaws of the single approaches.

\section{Experimental Results}
\label{sec:results}
To get an impression of how well the tool is working and to be able to research for better extraction methods, quality checks have been implemented.
Two test cases have been constructed, in cooperation with the two industry partners, to determine the performance of the approach. Each test case is using a policy document and a set of organisational metrics for which the evidence has been annotated in the document. The two test cases differ mainly in the length of the policy document considered (14 vs 69 pages).

The annotations were edited using the INCEpTION tool, as explained in Section~\ref{subsec:policy_docs}. A numerical analysis and a manual analysis were conducted for each approach. In the following, the two test cases are described and the results of the experiments are reported.

The first test case consists of a short document containing policies and 28 organisational metrics have been annotated for this document. The document belonging to the second test case is longer than the previous one and 50 organisational metrics have been annotated in it.

The annotations of the short document have been made by an expert person with in-depth knowledge of the context, while the longer document has been annotated by people in the field of cloud computing, but with only high-level knowledge of the context.

\subsection{Computed results}
\label{subsec:computed_results}
The various approaches presented in Section~\ref{sec:evidence_extraction} have been evaluated and compared by computing a \textit{quality} score that is defined as in Equation~\ref{eq:eq3}. 

\begin{equation}\label{eq:eq3}
score=\frac{\#correctly\ retrieved\ evidence}{\#total\ annotated\ evidence}
\end{equation}

This value represents the ratio between the number of correctly retrieved answers and the total number of annotated answers. A piece of evidence is counted as correctly retrieved if the tokens of the retrieved answer intersect with the annotated tokens. Tokens are chunks of characters from a text and are separated by spaces or punctuation marks. The results for the test cases can be seen in Table~\ref{tab:tab_res}.

\begin{table}[bht!]
\begin{center}

\setlength{\tabcolsep}{0.5em} 
\caption{Results obtained for each approach on the test policy documents}\label{tab:tab_res}
\begin{tabular}{@{}lrr@{}}
\toprule
\textbf{Approach} & \textbf{ShortDoc Score} & \textbf{LongDoc Score}\\
\midrule
Whole-doc               & 0.54 &  0.08\\
Keyword-based           & 0.68 &  0.26\\
Score-based             & 0.46 &  0.10\\
Similarity-based        & 0.25 &  0.12\\
Similarity + score-based  & 0.46 &  0.16\\

\bottomrule
\end{tabular}

\end{center}
\end{table}

The results show a clear deterioration when the long document is considered instead of the short one. This is because, for the short document, annotations are more specific and less ambiguous. Moreover, performance can suffer from the annotation bias introduced by non-expert annotators.

Further issues can derive from the particular choice of metric keywords included in each metric. Nevertheless, the keyword-based approach is the one that achieves the best performance for both test cases. On the other hand, the similarity-based and the similarity+score-based approaches seem not to benefit from the search space reduction based on keywords.

In general, the results obtained for the long document are limited and the division of the text into sections does not seem to lead to a particular improvement. This may also be due to inaccurate detection of sections or a poor structure of the document.

\subsection{Empirical results}
\label{subsec:empirical_results}
As seen from the results obtained, the analysis of the long document is the most problematic. For this reason, an empirical analysis on the long document has been performed, by inspecting the answers retrieved. To this aim, a set of 118 metrics have been evaluated and assessed. Among these metrics, 82 entries were manually set to \textit{compliant} and 36 were set to \textit{not compliant}. In addition, a comment has been provided by the evaluator to describe why the result is compliant or not. According to the information found in the comment, the metrics have been then grouped into 4 categories:
\begin{itemize}
    \item \textit{Correct}: this category includes all the results assessed as compliant and no comment has been provided.
    \item \textit{Partial matching}: it includes all results that have been marked as compliant but had additional comments indicating that the actual result would be in a different place. In any case, the systems' answer was correct. 
    \item \textit{False/other error}: it includes all results that have been set to not compliant and/or errors have been indicated within the comment. 
    \item \textit{Not in document}:  there are all results that had a comment indicating that the metric target is not contained in the policy document.
\end{itemize}

Table~\ref{tab_emp_res} reports the results obtained by applying the manual assessment. In the first column, the 4 error categories are listed, while the second and the third ones report the number of results belonging to that category and the percentage with respect to the whole results, respectively.

\begin{table}[bht!]
\begin{center}

\caption{Results from manual assessment. Each extracted evidence has been assessed and assigned an \textit{error category}. The keyword-based approach was used for evidence extraction.}\label{tab_emp_res}
\begin{tabular}{lrr}
\toprule
\textbf{Error category}  & \textbf{Count} & \textbf{Percentage} \\
\midrule
No error & 68 & 57.63\%\\
Partial matching   & 11 & 9.32\%\\
False/other error &  8 & 6.78\%\\
Not in document &  31 & 26.27\% \\
\midrule
\textbf{Total} &  118 & 100\%\\
\bottomrule
\end{tabular}

\end{center}
\end{table}

As a consequence of the above, there are at least three reasons why a result can be not compliant:
\begin{enumerate}
    \item The evidence is not in the document.
    \item The wrong answer was retrieved.
    \item The retrieved answer does not comply with the target value. 
\end{enumerate}

The first type of situation (evidence not in the document) is the one occurring most often, but in reality, this should not be considered an error, since the metric sought is not really present in the text. This problem can be solved by applying a method that is able to automatically detect whether or not the answer is present in the text, or by defining a set of metrics that are definitely present in the text. For example, the set of metrics could be defined by the staff responsible for policy creation, as they know which requirements/metrics are covered by the document.

For a quality measurement of the actual evidence retrieval system, the metrics considered \textit{not in document} have been excluded and the \textit{partial matching} have been counted as false results. This leads to an accuracy of 68/87=78.16\% (67 correctly retrieved answers vs 19 that are either only partly correct or not at all). 

\section{Conclusion and Future Work}
\label{sec:conclusion}

In this paper, we presented \ac{AMOE}, a tool to process organisational evidence to enable computer-aided audits. Several approaches have been proposed, to facilitate the automatic assessment of organizational evidence, such as the one available in security policy documents of \ac{CSP}s. This tool is also expected to reduce the time spent by auditors to check policy documents to filter and extract relevant information. The proposed approaches have been tested on an annotated dataset, providing promising results. The empirical results indicated that the correct answer could be retrieved for more than half of the tested metrics. In the future, these approaches need to be verified on larger sets of data and refined taking into account the context. Currently, the final decision on the assessment is still left to a human auditor, but once the system has been tested on a larger number of data and refined, the audit may be fully automated and only random samples need to be checked once in a while.

Furthermore,  additional \ac{QA} models could be used, and extractive text summarization techniques can be also applied. Generative text models could also be tested, e.g. those based on Generative Pre-trained Transformer-3 (GPT-3), but despite their powerfulness, they do not seem to be suitable for this task since they suffer uncertainty and correctness issues. Moreover, incorporating context in generative models can increase the computational cost and time. 

%
%
%
\bibliographystyle{splncs04}
\bibliography{resources}

\end{document}